\documentclass[a4,11pt]{article}
\usepackage{hyperref}
\usepackage{amsmath}  

\usepackage{url}

\usepackage{amssymb}  
\usepackage{float} 
\usepackage{xcolor} 
\usepackage{graphicx} 
\usepackage{cite} 
\usepackage[normalem]{ulem}
\usepackage{bbm}
\usepackage{dsfont}
\usepackage[mathscr]{euscript}
\usepackage{ntheorem}

\newcommand{\proof}{\noindent{\sc Proof:}}
\newcommand{\qed}{\hfill$\Box$ }

\newtheorem{theorem}{\sc  Theorem\rm}[section]

\newtheorem{lemma}[theorem]{\sc Lemma\rm}

\newtheorem{prop}[theorem]{\sc Proposition\rm}
\newtheorem{proposition}[theorem]{\sc Proposition\rm}

\newtheorem{remark}[theorem]{\sc Remark\rm}
\newtheorem{Remark}[theorem]{\sc Remark\rm}

\usepackage{bbm}
\usepackage{multirow}

\usepackage{hyperref}

\usepackage{cite}
\usepackage{graphicx}

\usepackage{color}
\usepackage{url}
\usepackage{wasysym}

\usepackage{mhequ}
\usepackage{psfrag}

\usepackage[normalem]{ulem}

\newcommand{\elltor}{\red{r}}

\newcommand{\V}{\red{\mathbb{V}}}
\newcommand{\Xtou}{\red{u}}
\newcommand{\Xtow}{\red{w}}
\newcommand{\Ytov}{\red{v}}

\newcommand{\Ztoz}{\red{z}}

\newcommand{\redc}{\color{red}}

\newcommand{\waehr}[1]{#1} 
\newcommand{\fuerErg}[1]{} 



\makeatletter
\newcommand{\pushright}[1]{\ifmeasuring@#1\else\omit\hfill$\displaystyle#1$\fi\ignorespaces}
\newcommand{\pushleft}[1]{\ifmeasuring@#1\else\omit$\displaystyle#1$\hfill\fi\ignorespaces}
\makeatother

\newcommand{\copright}[1]{\mnote{copyright: #1}}
\newcommand{\cpright}[1]{\copright{#1}}
\renewcommand{\cpright}[1]{\mnote{cpright}}
\newcommand{\cprights}[1]{\mnote{cpright settled}}
\newcommand{\cprightm}[1]{}
\newcommand{\reread}[1]{}

{\catcode `\@=11 \global\let\AddToReset=\@addtoreset}
\AddToReset{equation}{section}

\newcommand{\noViennaLectures}[1]{}

\newcommand{\red}[1]{{{\color{red}{#1}}}}

\newcommand{\qedskip}{\qed\bigskip}

\newcommand{\olstop}[3]{}
\newcommand{\lstop}[3]{}

\newcommand{\beijing}[1]{}

\newcommand{\beq}{\begin{equation}}

\newcommand{\FS}       
                  {F}

\newcommand{\HS} 
       {H_{\mbox{\scriptsize volume}}}

{\ptc{this should be removed in the oberwolfach version}}%

\newcommand{\eeal}[1]{\label{#1}\end{eqnarray}}
\newcommand{\bed}{\begin{deqarr}}
\newcommand{\eed}{\end{deqarr}}
\newcommand{\bedl}[1]{\begin{deqarr}\label{#1}}
\newcommand{\eedl}[2]{\arrlabel{#1}\label{#2}\end{deqarr}}

\newcommand{\bel}[1]{\begin{equation}\label{#1}}
\newcommand{\bea}{\begin{eqnarray}}
\newcommand{\bean}{\begin{eqnarray}\nonumber}
\newcommand{\beal}[1]{\begin{eqnarray}\label{#1}}
\newcommand{\eea}{\end{eqnarray}}

\newcommand{\be}{\begin{equation}}
\newcommand{\eeq}{\end{equation}}
\newcommand{\ee}{\end{equation}}
\newcommand{\beqa}{\begin{eqnarray}}
\newcommand{\eeqa}{\end{eqnarray}}
\newcommand{\beqan}{\begin{eqnarray*}}
\newcommand{\eeqan}{\end{eqnarray*}}
\newcommand{\ba}{\begin{array}}
\newcommand{\ea}{\end{array}}

\newcommand{\Id}{\mbox{\rm Id}} 

\DeclareFontFamily{OT1}{rsfs}{}
\DeclareFontShape{OT1}{rsfs}{m}{n}{ <-7> rsfs5 <7-10> rsfs7 <10-> rsfs10}{}
\DeclareMathAlphabet{\mycal}{OT1}{rsfs}{m}{n}

\newcounter{mnotecount}[section]

\renewcommand{\themnotecount}{\thesection.\arabic{mnotecount}}

\newcommand{\mnotex}[1]
{\protect{\stepcounter{mnotecount}}$^{\mbox{\footnotesize
$
\bullet$\themnotecount}}$ \marginpar{
\raggedright\tiny\em
$\!\!\!\!\!\!\,\bullet$\themnotecount: #1} }

\newcommand{\ptc}[1]{\mnote{{\bf ptc:}#1}}

\newcommand{\mnote}[1]
{\protect{\stepcounter{mnotecount}}$^{\mbox{\footnotesize
$
\bullet$\themnotecount}}$ \marginpar{
\raggedright\tiny\em
$\!\!\!\!\!\!\,\bullet$\themnotecount: #1} }

\newcommand{\warn}[1]
{\protect{\stepcounter{mnotecount}}$^{\mbox{\footnotesize
$
\bullet$\themnotecount}}$ \marginpar{
\raggedright\tiny\em
$\!\!\!\!\!\!\,\bullet$\themnotecount: {\bf Warning:} #1} }

\newcommand{\R}{\mathbb R}

\newcommand{\N}{\mathbb N}

\newcommand{\mcC}{{\mycal C}}

\newcommand{\beqar}{\begin{deqarr}}
\newcommand{\eeqar}{\end{deqarr}}

\newcommand{\beaa}{\begin{eqnarray*}}
\newcommand{\eeaa}{\end{eqnarray*}}

\newcommand{\tr}{\mbox{tr}}

 \renewcommand{\red}[1]{{#1}}
 
\renewcommand{\redc}{}

\renewcommand{\waehr}[1]{}
\begin{document}
%
\title{Thomas-Wigner  rotation
via Clifford algebras
\footnote{Vienna preprint UWThPh 2025-1}
}
\author{Piotr T. Chru\'sciel and Helmuth Urbantke
\\ Faculty of Physics, University of Vienna}
\maketitle

\abstract{We derive  Macfarlane's formula for the Thomas-Wigner angle of rotation using Clifford-algebra methods. The presentation is elementary and pedagogical, suitable for students with some basic knowledge of special relativity; no prior knowledge of Clifford algebras is required.}
\tableofcontents

\section{Introduction}

The aim of these notes is to provide a self-sufficient and relatively concise derivation of the angle, as well as   the direction of the axis, of the Thomas-Wigner  rotation.
Our presentation essentially coincides with Section~III of \cite{UrbantkeThomas}, with  the calculations described there expanded to make them suitable for a few lectures at undergraduate level. The approach to this question in \cite{UrbantkeThomas} has the advantage of introducing Clifford algebras to  students, making clear their usefulness  in elementary special relativity.

We thus provide a relatively simple derivation, as far as algebraic manipulations are concerned, of the following fact:

\begin{quote}
 Consider the rest frame   associated with a four-velocity vector $u$  so that,  using units where the speed of light $c$ equals 1, we have
      \begin{equation}\label{4II22.01}
            u = \left(
              \begin{array}{c}
                1 \\
                \vec 0 \\
              \end{array}
            \right)
            \,.
      \end{equation}
    Let  $v$ and $w$ be two timelike vectors which  in  this frame take the form
      \begin{equation}\label{4II22.1}
        v = \gamma_{uv} \left(
              \begin{array}{c}
                1 \\
                \vec v \\
              \end{array}
            \right)
            \,,
            \quad
        w = \gamma_{uw}
         \left(
              \begin{array}{c}
                1\\
                \vec w \\
              \end{array}
            \right)
            \,,
      \end{equation}
      with
      \begin{eqnarray}
       &
       \displaystyle
        \gamma_{uv}
         =  -\eta(u,v) = \frac{1}{\sqrt{1- \vec v^2}}
         \,,
          \
        \gamma_{uw}
         =  -\eta(u,w) = \frac{1}{\sqrt{1- \vec w^2}}
         \,,
         \phantom{xxx}
      \end{eqnarray}
      where $\eta$ denotes the Minkowski scalar product.
      Then the Lorentz transformation, which consists of a boost mapping $u$ to $v$ and then $v$ to $w$, differs from a boost mapping $u$ to $w$ by a rotation, in the oriented plane spanned by $\vec w$ and $\vec v$,
       by an angle
        $\psi$ satisfying
      \begin{equation}\label{4II22.2}
 \cos (\psi) =
    \frac{(1    + \gamma_{uv} + \gamma_{uw} + \gamma_{vw})^2}{ ( 1 + \gamma_{uv})(1+\gamma_{uw})(1+\gamma_{vw}) } - 1
      \end{equation}
      where, in the frame in which \eqref{4II22.1} holds,
      \begin{eqnarray}
       &
       \displaystyle
        \gamma_{vw} =  - \eta(v,w) = \gamma_{uv}  \gamma_{uw} (1- \vec v \vec w)
         \,.
          \label{5II22.31}
      \end{eqnarray}
\end{quote}

Indeed, the aim of this note is to prove Formula~\eqref{4II22.2}, which is attributed to Macfarlane~\cite{Macfarlane} in \cite{UrbantkeThomas}. The proof can be found   in Section~\ref{s5II22.5}; the rest of the manuscript  introduces the tools needed for the proof.

\section{Conventions and definitions}

In what follows we consider a real vector space $\red{\V}$ equipped with a Lorentzian scalar product $\eta$.

We use the \emph{mostly plus} convention, in which the signature of  $\eta$  is $(-,+,\ldots,+)$. This choice of signature is very convenient when passing from special to general relativity, as then spacelike hypersurfaces inherit directly a Riemannian metric from the spacetime one. In particle physics the \emph{mostly minus} convention is more useful, as then the square $\eta(p,p)$ of a four-momentum vector $p$ is $m^2$ instead of $-m^2$.

We use the summation convention and for $u,v\in\V$ we write
$$
\eta(\Xtou ,\Ytov  ) =\eta_{\mu\nu}\Xtou ^\mu \Ytov  ^\nu\,,
 \qquad
 |\Xtou | = \sqrt{|\eta(\Xtou ,\Xtou )|}
 \,,
$$
where  $|u|$ is referred to as \emph{Lorentzian length} of $u$.
A vector $\Xtou $ is said to be \emph{null} if $\eta(\Xtou ,\Xtou )=0$ but $\Xtou \ne 0$;
it is spacelike if $\eta(u,u) >  0$ \emph{or}  $u=0$;   timelike if $\eta(u,u)<0$.
A vector   is called \emph{unit} if $|u|=1$.

 Let $\Omega$ be a subset of $\red{\V}$, then $\Omega^\perp$ is the collection  of those elements of $\red{\V}$ which are orthogonal to all elements of $\Omega$.
For example, if $u$ is a non-zero vector, then $u^\perp:=\{u\}^\perp$
is a vector space of codimension one, i.e.\ of one dimension less than the dimension of $\red{\V}$. If $u$ is timelike, which in our signature means that $\eta(u,u)<0$, then all elements of $u^\perp$ are spacelike.

A \emph{Lorentz transformation} $\Lambda$ is a linear transformation of $\red{\V}$ which leaves the scalar product invariant, i.e.
\begin{equation}\label{25I24}
  \eta(\Lambda u,\Lambda v) = \eta(u,v)
  \,.
\end{equation}
A Lorentz transformation is \emph{proper} if its determinant equals one; it is called \emph{ortochronous} if it preserves time orientation.

Consider two  timelike vectors $u$ and $v$ in $\red{\V}$ of same Lorentzian length. We say that $u$ and $v$ are \emph{consistently time oriented} if $\eta(u,v)<0$. (The minus sign is determined by our choice of signature.)  Given a consistently time-oriented pair of timelike vectors we  define a \emph{boost} $B(v,u)$ as that \underline{proper} Lorentz transformation which

\begin{enumerate}
  \item maps $u$ into $v$: $B(v,u)\, u = v$, and
  \item is the identity on vectors which are orthogonal both to $u$ and $v$:
  $$ B(v,u)|_{\mathrm{Span}\{u,v\}^\perp} = \mathrm{Id}
      \,.
      $$
\end{enumerate}
The reader might want to verify that the requirements spelled-out above define $B(u,v)$ uniquely.
This can e.g.\ be done by showing that every two-dimensional proper Lorentz transformation which preserves time-orientation can be written as in \eqref{10I25.2} below.

A subspace $W\subset \red{\V}$ is called \emph{timelike} if the restriction of the metric $\eta$   to $W$, which we denote by $\eta_W$, has Lorentzian signature. Then there exists a basis $\{e_0,e_1\}$ of $W$ so that
\begin{equation}\label{10I25.1}
  \left(
    \begin{array}{cc}
      \eta_W(e_0,e_0) & \eta_W(e_0,e_1) \\
      \eta_W(e_1,e_0)  & \eta_W(e_1,e_1) \\
    \end{array}
  \right)
  :=
   \left(
    \begin{array}{cc}
      \eta(e_0,e_0) & \eta(e_0,e_1) \\
      \eta(e_1,e_0)  & \eta(e_1,e_1) \\
    \end{array}
  \right)=
  \left(
    \begin{array}{cc}
      -1 & 0 \\
      0  & 1 \\
    \end{array}
  \right)
  \,.
\end{equation}
In this basis every
proper Lorentz transformation $\Lambda$ which preserves $W$ and which preserves time-orientation
can be written as
\begin{equation}\label{10I25.2}
 \Lambda = \Lambda(\alpha):=  \left(
    \begin{array}{cc}
      \cosh(\alpha) & \sinh(\alpha) \\
      \sinh(\alpha)  &  \cosh(\alpha)\\
    \end{array}
  \right)
  \,,
\end{equation}
for some $\alpha \in \R$.
The parameter $\alpha$ is called the \emph{rapidity}.

Note that the replacement  $e_1\mapsto -e_1$, which changes the orientation of the basis $\{e_0,e_1\}$,  reverses the sign of $\alpha$.
So, similarly to the angle of rotation of an oriented plane,  the sign of $\alpha$ depends upon the choice of orientation.

Recall that the trace of a matrix is basis-independent. Calculating the trace of the right-hand side of \eqref{10I25.2}, we see that a basis- and orientation-independent way of calculating $\cosh(\alpha)$ is to calculate the trace of $\Lambda$, viewed as a map from $W$ to $W$:
\begin{equation}\label{10I25.41}
  \tr \Lambda = 2 \cosh(\alpha)
  \,.
\end{equation}
A consequence of this is,  that the rapidity of a boost $\Lambda =B(u,v)$ can be read-off from the hyperbolic angle between $u$ and $v$, which is defined quite generally for pairs of timelike vectors as
\begin{equation}\label{13I25.1}
  \cosh(\alpha) := - \frac{\eta(u,v)}{\sqrt{\eta(u,u) \eta(v,v)}}
   \,.
\end{equation}
Indeed,   let us choose an orthonormal  basis $\{e_0,e_1\}$  of $\mathrm{Span}\{u,v\}$ so that $u$ is proportional to the first basis vector, $u= \sqrt{|\eta(u,u) |} e_0$. In this basis we have, using  \eqref{10I25.2},
\begin{align}
 v=
  & \
   \Lambda u = \sqrt{|\eta(u,u) |}  \left(
    \begin{array}{cc}
      \cosh(\alpha) & \sinh(\alpha) \\
      \sinh(\alpha)  &  \cosh(\alpha)\\
    \end{array}
  \right)  \left(
    \begin{array}{c}
      1 \\
      0
      \\
    \end{array}
  \right)
 \nonumber
  \\
 = & \  \sqrt{|\eta(u,u) |} ( \cosh(\alpha) e_0 + \sinh(\alpha)e_1)
  \,,
\end{align}
which implies, using   \eqref{10I25.1},
\begin{eqnarray}
 \eta(u,\Lambda u)
   &=&   \eta(u,v)
  \nonumber
  \\  &=&  \eta
 \Big(
  \sqrt{|\eta(u,u) |}   e_0
 ,\sqrt{|\eta(u,u) |}
  \big( \cosh(\alpha) e_0 + \sinh(\alpha)e_1
   \big)
 \Big)
  \nonumber
  \\
   &=&   |\eta(u,u) |\cosh(\alpha) \eta(e_0,e_0)
  \nonumber
  \\
   &=&   -|\eta(u,u) | \cosh(\alpha)
  \,,
   \label{17I25.61}
\end{eqnarray}
which is the same as \eqref{13I25.1} since the lengths of $u$ and $v$ are equal.

It is immediate from \eqref{10I25.2} that
\begin{equation}\label{10I25.3}
  \Lambda(\alpha)\Lambda(\beta)= \Lambda(\alpha+\beta)
  \,.
\end{equation}
This implies in particular that the definition of rapidity is independent of the choice of positively oriented bases.

A \emph{rotation} is a proper Lorentz transformation  $R$ which leaves a timelike vector $u$ invariant. We then say that $R$ is a rotation in the rest frame of $u$.

\subsection{Orientation}

Given  vectors $\Xtou ,\Ytov  \in \V $ we denote by
 $\mathrm{Span}\{\Xtou ,\Ytov  \}$ the \emph{oriented} two-dimensional vector space consisting of all linear combinations of $\Xtou $ and $\Ytov  $.
Here  $\{u,v\}$ denotes the set containing $u$ and $v$, not to be confused with the anti-commutator which will appear in some formulae below.

Recall that an orientation of $\red{\V}$ is a choice of a family of bases of $\red{\V}$ such that the transition matrix between the bases has positive determinant.
We choose the orientation of $\mathrm{Span}\{\Xtou ,\Ytov  \}$ by declaring that
$\{\Xtou ,\Ytov  \}$ is positively oriented.

As an example, let $\{e_1,e_2\}$ be a basis of $\V$, and suppose that in this basis we have
\begin{equation}\label{17I25.1}
  u = \left(
        \begin{array}{c}
          1 \\
          0 \\
        \end{array}
      \right)
      \,,
      \quad
  v = \left(
        \begin{array}{c}
          \cos(\theta) \\
          \sin(\theta)\\
        \end{array}
      \right)
      \, ,
\end{equation}
with $\theta\in(0,2\pi)\setminus \{\pi\}$; we have assumed $\theta\ne 0,\pi, 2\pi$ to avoid the cases  where $u$ and $v$ are colinear.
Then the basis $\{e_1,e_2\}$ will be positively oriented in  $\mathrm{Span}\{\Xtou ,\Ytov  \}$ if and only if the determinant of the matrix
\begin{equation}\label{17I25.2}
    \left(
        \begin{array}{cc}
          1 &
          \cos(\theta) \\
      0&    \sin(\theta)\\
        \end{array}
      \right)
\end{equation}
  is positive, i.e. if and only if $\theta\in(0,\pi)$.

\subsection{Angles}

Finally, some remarks on angles are in order.
Let $\langle \cdot,\cdot\rangle$ denote a (usual, i.e.\ positive-definite) scalar product on an $n$-dimensional vector space $\mathbb{V}$, with  $2\le n\in  \N\ \cup \{\infty\}$.
Given two non-zero vectors $u,v\in \mathbb{V}$,
the \emph{non-oriented angle}%
\footnote{The ``non-oriented'' terminology comes from~\cite{Berger}, where in addition to \emph{non-oriented angle} one defines an \emph{oriented angle} between $u$ and $v$  as the rotation which maps
$u/ \sqrt{\langle u,u\rangle}$ into
$v/ \sqrt{\langle v,v\rangle}$.}
$$
 \theta\in [0,\pi]
$$
between $u$ and $v$, or simply \emph{angle}, is defined as
\begin{equation}\label{1I25.1}
  \cos\theta =\frac{ \langle u,v\rangle}{\sqrt{\langle u, u \rangle
   \langle v, v \rangle }}
    \,.
\end{equation}

We finish these preliminaries by introducing the notion of \emph{angle of rotation of an oriented plane}, defined as follows:

 Let $(\V,\langle \cdot,\cdot\rangle)$ be an oriented $2$-dimensional vector space with  scalar product $\langle \cdot,\cdot\rangle$.
Let $\{e_1,e_2\}$ be a positively oriented basis  of $\V$. We define a \emph{rotation of angle $\theta$} as the map which, in this basis, is represented by the matrix
\begin{equation}\label{16I25.11}
  R(\theta):= \left(
     \begin{array}{cc}
       \cos(\theta) & -\sin(\theta) \\
       \sin(\theta) & \cos(\theta) \\
     \end{array}
   \right)
    \,.
\end{equation}
It is not very difficult to check, using the fact that
$\mathrm{SO}(2)$ is commutative, that the matrix representation \eqref{16I25.11} of $R(\theta)$  is independent of the positively-oriented basis chosen.

We allow any angle of rotation $\theta\in\R$, even though of course we have $R(\theta+2\pi)=R(\theta)$.

In this definition the orientation of $\V$ has been used, and will often be emphasised in what follows. To see why this is the case, suppose  that we change the basis $(e_1,e_2)$ to one which is oppositely oriented with respect  to the original one. For example, let $(f_1,f_2)= (-e_1,e_2)$,  then
\begin{eqnarray*}
  R(\theta)f_1
   & = &
    - R(\theta) e_1 = -
       \cos(\theta)e_1 -
       \sin(\theta)e_2 =
       \cos(\theta)f_1 -
       \sin(\theta)f_2 \,,
       \nonumber
\\
  R(\theta)f_2
   & = &
      R(\theta) e_2 =
      - \sin(\theta)e_1+
       \cos(\theta)e_2 =
        \sin(\theta)f_1+
       \cos(\theta)f_2 \,,
\end{eqnarray*}
In this basis the matrix of $R(\theta)$ therefore reads
\begin{equation}\label{16I25.12}
 \left(
     \begin{array}{cc}
       \cos(\theta) &  \sin(\theta) \\
      - \sin(\theta) & \cos(\theta) \\
     \end{array}
   \right) = \left(
     \begin{array}{cc}
       \cos(-\theta) &  -\sin(-\theta) \\
       \sin(- \theta) & \cos(-\theta) \\
     \end{array}
   \right)
    \,.
\end{equation}
This takes the same form as \eqref{16I25.11} with $\theta$ replaced by $-\theta$. This can be interpreted as a change of sign of the rotation parameter of $R(\theta)$. This ambiguity is removed when only positively-oriented bases are allowed.

Note that the angle \eqref{1I25.1} between $u$ and $v$ is the same as the angle between $v$ and $u$. But suppose that a rotation of angle $\theta$ maps $u$ to $v$, then one needs a rotation of angle $-\theta$ to map $v$ to $u$.

\section{Thomas-Wigner  rotation}

Consider three unit timelike vectors $u$, $v$ and $w$. We wish to analyse the Lorentz transformation obtained by composing a boost which maps $u$ to $v$, followed by a boost that maps $v$ to $w$, and followed by a boost that maps $w$ back to $u$;
\begin{equation}\label{29I22.3}
 R:= B(u,w) B(w,v) B(v,u)
\,.
\end{equation}
Silberstein~\cite{SilbersteinBook} (see also Thomas~\cite{ThomasPrecession} and Wigner~\cite{WignerPrecession})
 pointed out that $R$ is a nontrivial rotation in a spacelike hyperplane orthogonal to $u$. This is rather clear:
By definition $R$ is a Lorentz transformation that maps $u$ to $u$, and is proper, and is  thus  a rotation in the hyperplane orthogonal to $u$.

The map $R$ is often referred to as  the \emph{Thomas-Wigner  rotation}, and we will follow this terminology.

The point of what follows is to get insight into the structure of $B(v,u)$, in particular express it in tractable form.

Now, $B(v,u)$ is the identity on
$$\mathrm{Span}\{u,v\}^\perp =u^\perp \cap \red{v^\perp}
\,,
$$
and $B(w,v)$
is the identity on $\red{v^\perp} \cap \red{w^\perp}$, and $B(u,w)$
is the identity on $\red{w^\perp} \cap u^\perp$. It follows that $R$ is the identity on
$$
 u^\perp \cap \red{v^\perp}  \cap \red{w^\perp} = \mathrm{Span}\{u,v,w\}^\perp
 \,,
$$
which is a subspace of codimension  three whenever $u$, $v$ and $w$ are linearly independent. It follows that the interesting action takes place in the three dimensional space $ \mathrm{Span}\{u,v,w\}$, which is Lorentzian whenever one of the vectors $u$, $v$ and $w$ is timelike. So while the analysis below applies in all dimensions, it would in fact suffice  to consider a three-dimensional Lorentzian vector space to obtain our main conclusions.

\section{Reflections}
 \label{s13I25.1}

Given a  vector $\Xtou $ such that $\eta(\Xtou ,\Xtou )\ne0$ (thus, $\Xtou \ne 0$ and $\Xtou $ is \emph{not} null), consider the linear map $S_\Xtou $ from $\red{\V}$ to $\red{\V}$ defined as
\begin{equation}\label{30I22.5}
 S_\Xtou (\Ytov  ) =  \Ytov   -  2 \frac{\eta(\Xtou ,\Ytov  )  }{\eta(\Xtou ,\Xtou ) }\Xtou
 \,.
\end{equation}

A useful property of $S_\Xtou $ is that it does not depend upon rescalings of $\Xtou $: if $\lambda \in \mathbbm{R}^*$ then
\begin{eqnarray}
 \nonumber 
 S_{\lambda \Xtou } (\Ytov  )
   &=&
   \Ytov   -  2 \frac{\eta(\lambda \Xtou ,\Ytov  )  }{\eta(\lambda \Xtou ,\lambda \Xtou ) }\lambda \Xtou  =   \Ytov    -   2 \frac{\eta(\Xtou ,\Ytov  )  }{\eta(\Xtou ,\Xtou ) }\Xtou
    \nonumber
\\
 & = & S_\Xtou  (\Ytov  )
 \,.
  \label{29I22.11a}
\end{eqnarray}

We claim that $S_\Xtou $ is a Lorentz transformation which is an \emph{orthogonal reflection with respect to the plane orthogonal to $\Xtou $}. For this, we start by noting that, for calculational purposes, it is convenient to make use of the scaling freedom \eqref{29I22.11a}, which implies that it suffices to carry out the argument  for vectors $\Xtou $ satisfying
\begin{equation}\label{28I22.6}
  \varepsilon:= \eta(\Xtou ,\Xtou ) \in \{\pm 1\}
  \,.
\end{equation}
One then has the less-cluttered formula
\begin{equation}\label{28I22.7}
  S_\Xtou  (\Ytov   ) = \Ytov    -  2 \varepsilon \eta(\Xtou ,\Ytov  ) \Xtou
  \,,
\end{equation}
with $\varepsilon^2=1$.

Now, clearly, $S_\Xtou $ is a linear map which is the identity on vectors orthogonal to $\Xtou $, with
$$
 S_\Xtou (\Xtou )=-\Xtou
  \,.
$$
This already implies, with a little thought, that $S_\Xtou $ is a Lorentz transformation.
Alternatively, this is   a   two-lines exercice which uses linearity and symmetry of the  scalar product:
\begin{eqnarray}
  \eta\big(  S_\Xtou   (\Ytov  ),   S_\Xtou   (\Ztoz ) \big)  & = &
    \eta\big(  \Ytov   -  2 \varepsilon \eta(\Xtou ,\Ytov  ) \Xtou
    ,  \Ztoz  -  2 \varepsilon \eta(\Xtou ,\Ztoz ) \Xtou
    \big)
 \nonumber
\\
 & = &    \eta\big(   \Ytov  ,\Ztoz   \big)
    -  2 \varepsilon \eta(\Xtou ,\Ztoz )\eta\big(   \Ytov  , \Xtou   \big)
 \nonumber
\\
 &  &
 -      2 \varepsilon \eta(\Xtou ,\Ytov  )
 \eta\big( \Xtou
    ,  \Ztoz
    \big)  + 4  \underbrace{\varepsilon^2}_1  \eta(\Xtou ,\Ytov   ) \eta(\Xtou ,\Ztoz  )
 \underbrace{ \eta\big( \Xtou
    ,   \Xtou
    \big)}_{\varepsilon}
 \nonumber
\\
 & = &   \eta(\Ytov  ,\Ztoz )
 \,.
    \label{28I22.3xza}
\end{eqnarray}
So $\eta\big(  S_\Xtou   (\Ytov  ),   S_\Xtou   (\Ztoz ) \big) = \eta(\Ytov  ,\Ztoz )$, which is precisely the defining property of Lorentz transformations.

We also note that
\begin{eqnarray}
  S_\Xtou ^2(\Ytov  )
    & =  & S_\Xtou
     \big(
        \Ytov   -  2 \varepsilon \eta(\Xtou ,\Ytov  ) \Xtou
         \big)
   = S_\Xtou (  \Ytov  )  -  2 \varepsilon \eta(\Xtou ,\Ytov  )
    \underbrace{S_\Xtou (  \Xtou  )}_{-\Xtou }
     \nonumber
\\
 & = & \Ytov
  \,,
    \label{28I22.2}
\end{eqnarray}
thus $S_\Xtou ^2$ is the identity map, as appropriate for a reflection.

Finally, if $\Lambda$ is a Lorentz transformation we have $\eta(\Lambda \Xtou ,\Lambda \Xtou ) = \eta(\Xtou ,\Xtou )=\varepsilon$,  which can be used in the calculation that follows to show that
\begin{eqnarray}
  (\Lambda S_\Xtou  \Lambda^{-1})(\Ytov  )  & = &  \Lambda
   \big(
     \Lambda^{-1}\Ytov   -  2 \frac{\eta(\Xtou , \Lambda^{-1}\Ytov  )  }{\eta(\Xtou ,\Xtou ) }\Xtou
      \big)
 \nonumber
\\
 & = &       \Ytov   -  2 \varepsilon{\eta( \Lambda  \Xtou ,\Lambda \Lambda^{-1} \Ytov  )  }  \Lambda \Xtou
 \nonumber
\\
 & = &
  \Ytov   -  2 \varepsilon{\eta( \Lambda  \Xtou ,  \Ytov  )  }  \Lambda \Xtou
 \nonumber
\\
 & = &    S_{\Lambda \Xtou } \Ytov
 \,.
    \label{28I22.3}
\end{eqnarray}

Summarising, it holds that:
\begin{prop} For any non-null and non-zero vector $\Xtou $,
$S_\Xtou $ is a Lorentz transformation satisfying
\begin{equation}\label{28I22.9}
  S_\Xtou ^2=\mathrm{Id}
  \,,
  \quad
  S_\Xtou (\Xtou )= -\Xtou
  \,,
  \quad
  S_\Xtou |_{\Xtou ^\perp}=  \mathrm{Id}
  \,.
\end{equation}
Moreover for any Lorentz transformation $\Lambda$ we have
\begin{equation}\label{28I22.10}
  \Lambda  S_\Xtou  \Lambda ^{-1} = S_{\Lambda \Xtou }
  \,.
\end{equation}
Finally, if $\lambda$ is a non-zero real number it holds that
\begin{equation}\label{28I22.10a}
  S_{\lambda \Xtou }  = S_{ \Xtou }
  \,.
\end{equation}
\end{prop}

Some key facts for our further purposes are:

\begin{prop}
 \label{p28I22.2}
 \begin{enumerate}
   \item  If $a$ and $b$ are both not null and not zero, then $S_{b}S_{a}$ maps  $\mathrm{Span}\{a,b\} $ to itself, and is the identity on $\mathrm{Span}\{a,b\}^\perp$.

   \item
If $u$ and $v$ are both timelike and
consistently time oriented,
 with same lengths,
 then
\begin{equation}\label{28I22.11}
  S_{u+v}S_u =  B(v,u)
  \,.
\end{equation}
\item
Consider two linearly-independent vectors $a$ and $b$  which are   both orthogonal to a timelike vector $u$.
Let $\theta$ be the (non-oriented) angle $\theta\in (0,\pi)$
between them; by definition
\begin{equation}\label{13I51.1}
 \cos(\theta) = \frac{\eta(a,b)}{\sqrt{\eta(a,a)\eta(b,b)}}
\,.
\end{equation}
Then $S_{b}S_{a}$
  is a rotation by the angle  $2\theta\in(0,2\pi)$
 in the
 oriented
 spacelike plane $\mathrm{Span}\{a,b\}$.

\item
Consider two  distinct,  future directed, timelike vectors $u$ and $v$ of same Lorentzian length.
Let
$\alpha>0$
be the hyperbolic angle between them, defined for general timelike vectors as
$$
 \cosh(\alpha) =-  \frac{\eta(u,v)}{\sqrt{\eta(u,u)\eta(v,v )}}
\,.
$$
Then $S_{v}S_{u}$ is a boost with rapidity  $2\alpha$
 in the
 timelike plane $\mathrm{Span}\{u,v\}$.
\end{enumerate}
\end{prop}

\proof\
1.
If $\Xtou $ is a linear combination of $a$ and $b$, it is immediate from the definition \eqref{30I22.5} that both $S_a(\Xtou )$ and $S_b(\Xtou )$ are. Thus $S_b S_a$ maps $\mathrm{Span}\{a,b\}$ to itself.

Since  $ S_a$ is the identity on $a^\perp$  and $S_{b}$ is the identity on $b^\perp$, the map $S_{b}S_a$ is the identity on
$$
 a^\perp \cap b^\perp =   \mathrm{Span}\{a,b\}^\perp
  \,.
$$
2.
By point 1.\ the map  $S_{u+v}S_u $ is the identity on
$$
\mathrm{Span}\{u+v,u\}^\perp
 =
  \mathrm{Span}\{v,u\}^\perp = \mathrm{Span}\{u,v\}^\perp
  \,.
$$
Hence $S_{u+v}S_u $ is a proper ortochronous Lorentz transformation in the timelike plane $\mathrm{Span}\{u,v\}$.

It remains to show that the image of $u$ by $S_{u+v}S_u $  is $v$.
This is obtained by the following calculations, which apply regardless of the causal character of $u$ and $v$ as long as $u+v$ is \emph{not} null and \emph{not} zero, and that $u$ and $v$ have the same Lorentzian length: First,
\begin{eqnarray}
 \nonumber
   ( S_{u+v}S_u)( u)   &=&
    S_{u+v}\big( S_u ( u)\big)  = S_{u+v} (-u)
     =- S_{u+v} (u)
\\
  &  = &
   -
   \big( u  -2 \frac{\eta(u+v,u)}{\eta(u+v,u+v)} (u+v)
   \big)
    \label{29II22.1}
   \,.
\end{eqnarray}
Next, using that $\eta(u,u)=\eta(v,v)$,
\begin{eqnarray}
 \nonumber
  \frac{\eta(u+v,u)}{\eta(u+v,u+v)}   &=&
  \frac{\eta(u ,u)+\eta( v,u)}{\eta(u ,u)+\eta( v,u)+\eta(u,v )+\eta( v, v)}
\\
  &  = &
  \frac{\eta(u ,u) +\eta( v,u)}{ 2\eta(u ,u) +2\, \eta( v,u) } = \frac 12
   \,,
\end{eqnarray}
and \eqref{29II22.1} gives
\begin{eqnarray}
   ( S_{u+v}S_u)( u)   &=& v
    \label{29II22.2}
   \,.
\end{eqnarray}
which finishes the proof of point 2.

For further use we note that the last equation, together with $S_u(u)=-u$, implies
\begin{equation}\label{9I25.1}
     S_{u+v} (\underbrace{ u}_{-S(u)})  = - (S_{u+v}S_u) ( u) =-  v
     \,.
\end{equation}

\medskip

 3.
By point 1.\ the Lorentz transformation $S_\red{b} S_\red{a}$ is the identity on   $\mathrm{Span}\{\red{a,b}\}^\perp$, and leaves  $\mathrm{Span}\{\red{a,b}\}$ invariant. Since both $a$ and $b$ are orthogonal to a timelike vector, the plane $\mathrm{Span}\{\red{a,b}\}$ is spacelike.
 The statement does not depend upon the length of $a$ and $b$, so without loss of generality we can assume that both vectors  have the same length.
Further the sum  $a+b$ of two spacelike vectors $a$ and $b$, both orthogonal to a timelike vector, is spacelike and therefore cannot be null, and cannot vanish by the assumed linear independence, so that \eqref{9I25.1} applies:
\begin{eqnarray}
   ( S_{a+b}S_a)( a)   &=& b
    \label{29II22.2de}
   \,.
\end{eqnarray}
It follows that $S_{a+b}S_a$ is a rotation by some angle, the value of which must clearly be equal to $\theta$ given by \eqref{13I51.1}.

We have:
\begin{align}
  (S_{a+b}S_a)^2
   = &\
   S_{a+b}S_a \underbrace{S_{a+b}}_{(S_{a+b})^{-1}}S_a  & \mbox{by \eqref{28I22.9}}
    \nonumber \\
       =
       &\
   \underbrace{S_{a+b}S_a (S_{a+b})^{-1}}_{S_{S_{a+b}(a)}}S_a  & \mbox{by \eqref{28I22.10}}
   \nonumber
   \\
       =
       &\
  S_{
        {\underbrace{S_{a+b}(a)}_{-b}}
        }S_a  & \mbox{by \eqref{9I25.1}}
   \nonumber
   \\
       = &\
       \underbrace{ {S_{-b}}}_{
 S_{b} }
  S_a  & \mbox{by \eqref{28I22.10a}}
   \nonumber
   \\
       = &\
 S_{ b}S_{a}
   \,.
    &
   \label{8I25.1}
\end{align}
Thus $S_bS_a$ is the square of $S_{a+b}S_a$, which is a  rotation of the oriented plane $\mathrm{Span}\{a,b\}$ by an angle $\theta$, hence $S_bS_a$ is a rotation of
the oriented plane $\mathrm{Span}\{a,b\}$
by the angle $2\theta$.


4. From what has been said the map $S_vS_u =B(u,v)$ is a proper Lorentz transformation mapping  $\mathrm{Span}\{\red{u,v}\}$ to itself.  Without loss of generality we can assume that both $u$ and $v$ have unit length. It follows from~\eqref{13I25.1}-\eqref{17I25.61}  that the   rapidity of $S_vS_u $ can be read from the scalar product $-\eta(S_v S_u u ,u) $. We have
\begin{eqnarray}
  -\eta(S_v \underbrace{S_u u}_{-u},u)  &=& \eta(\underbrace{S_v u}_{u + 2\, \eta(u,v)v},u)
  \nonumber
  \\
    &=& \eta(u,u) +
     2\big( \eta(u,v)\big)^2
     \nonumber
      \\
    &=&
    -1  +2 \cosh^2(\alpha)
     \nonumber
     \\
    &=&
   \cosh(2\alpha)
    \,,
    \label{10I25.11}
\end{eqnarray}
which establishes the result.
%
%
\qedskip

\section{Clifford algebra}

A key tool for our derivation of the angle of rotation of the Thomas-Wigner  map $R$ defined in \eqref{29I22.3} will be the Clifford algebra associated with our vector space  $\red{\V}$ and its scalar product  $\eta$. As already mentioned, for our purposes it suffices to take $\V=\R^3$ with the Lorentzian scalar product, but note that there exists a Clifford algebra for any non-degenerate quadratic form on any finite dimensional vector space.

As such, the only thing we need to know is that there exists an associative real algebra $\mcC$, with unit element which we denote by  $\mathbbm{1} $, and with a linear embedding of our  vector space $\red{\V}$, say $\gamma: \V\to \mcC$, such that we have
\begin{equation}\label{28I22.12}
 \forall\  \Xtou ,\Ytov  \in \V \quad  \gamma(\Xtou ) \gamma(\Ytov  ) + \gamma (\Ytov  ) \gamma(\Xtou ) = 2\, \eta(\Xtou ,\Ytov  )
  \mathbbm{1}
 \,.
\end{equation}

If $\V=\R^3$ with the Lorentzian scalar product, and using a  basis $e_\mu$ for $\red{\V}$  so that $\eta(e_\mu,e_\nu)\equiv \eta_{\mu\nu}$ is a diagonal matrix with entries $(-1,+1,+1)$, the commutation relations \eqref{28I22.12} are usually written in the physics literature as
\begin{equation}\label{28I22.13}
   \gamma_\mu \gamma_\nu  + \gamma_\nu \gamma_\mu  = 2\, \eta_{\mu\nu}
   \,,
   \quad \mbox{where} \  \gamma_\mu:=\gamma(e_\mu)
 \,.
\end{equation}
In particular  the unit element $\mathbbm{1} $ at the right-hand side of \eqref{28I22.12} is omitted, which we will do from now on.

Equation~\eqref{28I22.13} is indeed the same as \eqref{28I22.12}, which can be verified by using linearity of $\gamma$ to write
\begin{eqnarray}
  &
  \gamma(\Xtou ) \gamma(\Ytov  ) + \gamma (\Ytov  ) \gamma(\Xtou )
   =\Xtou ^\mu \Ytov  ^\mu (\gamma_\mu \gamma_\nu  + \gamma_\nu \gamma_\mu)
  \,.
  &
\end{eqnarray}
Then the left-hand side will be equal to $2\eta(\Xtou ,\Ytov  )$ for all vectors $\Xtou $ and $\Ytov  $ if and only if \eqref{28I22.13} holds.

An example of a set of matrices $\gamma_\mu$ satisfying the above is given in Appendix~\ref{App30I22.1}.
Fortunately, such explicit matrix representations are never needed neither  here, nor for most calculations involving Clifford algebras.

It is convenient to forget the embedding map $\gamma$ in \eqref{28I22.12} (as well as the unit element $\mathbbm{1} $) and write instead
\begin{equation}\label{28I22.15}
\Xtou  \cdot \Ytov    +\Ytov   \cdot \Xtou  = 2\, \eta(\Xtou ,\Ytov  )
 \,.
\end{equation}
Note that some authors use the symbol $\Xtou \cdot \Ytov  $ to denote the scalar product of $\Xtou $ and $\Ytov  $. We emphasise that in this work ``$\cdot$'' denotes the Clifford algebra product $\gamma(\Xtou )\gamma(\Ytov  )$, with the scalar product written as $\eta(\Xtou ,\Ytov  )$ unless explicitly indicated otherwise.

The following observations are useful for calculations:
First, if $\Xtou $ and $\Ytov  $ are parallel or anti-parallel with  $\Xtou \ne 0$ we can write $\Ytov  =\lambda \Xtou $,
then
$$
 \Xtou  \cdot \Ytov   = \Xtou  \cdot (\lambda \Xtou ) =  \lambda \Xtou  \cdot \Xtou  = \Ytov   \cdot \Xtou
 \quad
 \Longrightarrow
 \quad
 [\Xtou ,\Ytov  ]=0
 \,.
$$
When  $\Xtou =0$ the commutator vanishes trivially since then both $\Xtou \cdot \Ytov  $ and $\Ytov  \cdot \Xtou $ are zero.
Thus,  collinear vectors commute, and we then have
\begin{equation}\label{7II22.3}
 \Xtou  \cdot \Ytov    + \Ytov   \cdot \Xtou  = 2 \Xtou  \cdot \Ytov   = 2\eta(\Xtou ,\Ytov  )
 \,,
\end{equation}
in particular
\begin{equation}\label{7II22.4}
 \Xtou  \cdot \Xtou =  \eta(\Xtou ,\Xtou )
 \,.
\end{equation}
Note that if  $\eta(u,u)\ne 0$, (equivalently, $u$ is not zero and not null) then $u$ is invertible with
\begin{equation}\label{10I25.31}
  u^{-1} = \frac{u}{\eta(u,u)}
   \,.
\end{equation}

Next, if $\Xtou $ and $\Ytov  $ are orthogonal, i.e.\ $\eta(\Xtou ,\Ytov  )=0$, then
$$
 0 = 2\, \eta(\Xtou ,\Ytov  ) = \underbrace{  \Xtou  \cdot \Ytov    +  \Ytov   \cdot \Xtou
 }_{=:
  \{\Xtou ,\Ytov  \}}
  \quad
  \Longleftrightarrow
  \quad
  \Xtou  \cdot \Ytov   = - \Ytov   \cdot \Xtou
 \,,
$$
where $\{\Xtou ,\Ytov  \}$ is called the \emph{anti-commutator} of $\Xtou $ and $\Ytov  $. Thus, $\Xtou $ and $\Ytov  $ anticommute if and only if they are orthogonal.

The above has another  corollary which is again useful as far as calculations are concerned. Namely, consider a vector $\Xtou \ne 0 $ which is  \emph{not null}. In such a case we can decompose any vector $\Ytov  $ as
\begin{equation}
 \label{2II22.5}
 \Ytov  =\red{\Ytov_\parallel}+\red{\Ytov_\perp}
 \,,
\ee
where $\red{\Ytov_\parallel} $ is parallel to $\Xtou $ and $ \red{\Ytov_\perp}$ is orthogonal to $\Xtou $:
\begin{equation}\label{2II21.1}
  \red{\Ytov_\parallel} := \frac{\eta(\Xtou ,\Ytov  )}{\eta(\Xtou ,\Xtou )} \Xtou
   \,,
   \qquad
    \red{\Ytov_\perp} := \Ytov   - \red{\Ytov_\parallel} = \Ytov  -   \frac{\eta(\Xtou ,\Ytov  )}{\eta(\Xtou ,\Xtou )} \Xtou
  \,.
\end{equation}
From what has been said it follows that
\begin{equation}\label{7II22.2}
  \Xtou \cdot \red{\Ytov_\parallel} = \red{\Ytov_\parallel}\cdot  \Xtou  = \eta(\Xtou ,\Ytov  )
  \,,
  \qquad
  \Xtou \cdot \red{\Ytov_\perp} = - \red{\Ytov_\perp} \cdot \Xtou
  \,.
\end{equation}
These formulae will be very useful shortly.

We emphasise that null vectors play a distinguished role here, as they are orthogonal to themselves. For such vectors we have
\begin{equation}\label{7II22.3asd}
  \Xtou \cdot \Xtou  = \eta(\Xtou ,\Xtou ) = 0 = - \Xtou  \cdot \Xtou
  \quad
   \Longleftrightarrow
   \quad
  [\Xtou ,\Xtou ] = \{\Xtou ,\Xtou \} = 0
  \,.
\end{equation}
Thus  a null vector provides  a non-zero element of the Clifford algebra which squares to zero, hence both commutes and anti-commutes with itself.

The above can \emph{almost} be summarised as follows
(what is missing so far is the proof that $[\Xtou ,\Ytov  ]=0$ implies that $\Xtou $ and $\Ytov  $ are collinear, which is irrelevant for what follows, and which we provide for completeness in Appendix~\ref{App7II22.1}):

\begin{prop}
 \label{P2II22.1}
 \begin{enumerate}
   \item
 Two vectors $\Xtou $ and $\Ytov  $ commute if and only if they are collinear.
   \item Two vectors $\Xtou $ and $\Ytov  $ anticommute if and only if they are orthogonal.
 \end{enumerate}
\end{prop}

The key fact that we will need for our analysis of the Thomas-Wigner  rotation matrix is:

\begin{prop}
Suppose that $\eta(u,u)\ne 0$,
 then
 \label{p28I22.3}
\begin{equation}\label{28I22.16}
 \forall \ \Ytov  \ \in \V
 \qquad
 \gamma(u)\gamma(\Ytov  )\gamma(u)^{-1} = - \gamma( S_u(\Ytov ))
 \,,
 \end{equation}
 where $S_u$ has been defined by \eqref{30I22.5}.
\end{prop}

In the notation of \eqref{28I22.15}, Equation \eqref{28I22.16} takes the form
\begin{equation}\label{28I22.16b}
 \forall \ \Ytov    \in \V
 \qquad
 u \cdot \Ytov   \cdot u ^{-1} =  - S_u(\Ytov)
 \,,
 \end{equation}
and this notation will be used from now on.
We  say that $-S_u(\Ytov )$ is the conjugate of $\Ytov  $ by $u$, and that $-S_u$ is the conjugation by $u$.

\medskip

{\noindent \sc Proof of Proposition \ref{p28I22.3}:}
Recall \eqref{10I25.31},
\begin{equation}\label{10I25.31c}
  u^{-1} = \frac{u}{\eta(u,u)}
   \,.
\end{equation}
Thus
\begin{eqnarray}
  u \cdot \Ytov   \cdot u^{-1} &=&   u \cdot \Ytov   \cdot \frac{u}{\eta(u,u)}
  =
      u \cdot    \frac{2\, \eta ( \Ytov   , u)   -  u \cdot \Ytov } {\eta(u,u)}
  \nonumber
    \\
    &=& \frac{ 2\, \eta(\Ytov  ,u)\, u  - u \cdot  u \cdot  \Ytov
      } {\eta(u,u)}
      =   2\,\eta(\Ytov ,u) \frac{u } {\eta(u,u)} -   \Ytov
  \nonumber
    \\
    &=& - S_u(\Ytov )
     \,,
\end{eqnarray}
as desired.
\qedskip

Now, in Section~\ref{s13I25.1} we used the definition \eqref{30I22.5}
to establish several properties of $S_u$. An alternative point of view is to   define $S_u$ by the left-hand side of \eqref{28I22.16b}. Let us take this approach now and rederive what has been said about $S_u$ using the Clifford algebra formalism.

So, suppose that $u$ is \emph{either} spacelike and non-zero \emph{or} timelike and set
\begin{equation}\label{28I22.16bc}
 S_u(\Ytov):= - u \cdot \Ytov   \cdot u ^{-1}
 \,.
 \end{equation}
Then
$$
S_u(u) =  - u \underbrace {\cdot  u   \cdot u ^{-1} }_{1}
 =-u
 \,,
$$
and so $S_u$ changes the sign of $u$.
Next, let $v$ be orthogonal to $u$. Then
$$
S_u(v)  = - \underbrace{  u\cdot v}_ { 2\eta(u,v)  - v \cdot u
 =- v \cdot u  } \cdot u^{-1}   = v   \cdot u \cdot u^{-1} = v
\,.
$$
We see that $S_u$ is the identity on $u^\perp$.
Further,
\begin{eqnarray}
  2\, \eta\big(S_u(\Ytov),S_u(w)\big)
     &=& S_u(\Ytov)\cdot  S_u(w)   +
  S_u(w) \cdot S_u(\Ytov)
   \nonumber
   \\
    &=& - u \cdot \Ytov   \cdot u ^{-1} \cdot ( - u \cdot w   \cdot u ^{-1} ) -   u \cdot w  \cdot u ^{-1} \cdot (-  u \cdot \Ytov   \cdot u ^{-1} )
   \nonumber
   \\
    &=&   u \cdot \Ytov    \cdot w   \cdot u ^{-1}   +   u \cdot w   \cdot \Ytov   \cdot u ^{-1}  \nonumber
   \\
    &=&   u \cdot
     (\Ytov    \cdot w     +     w   \cdot \Ytov )  \cdot u ^{-1}  \nonumber
   \\
    &=&   u \cdot (  2\, \eta\big(\Ytov,w\big)   ) \cdot u ^{-1}
    \nonumber
    \\
    &=&  2\, \eta\big(\Ytov,w\big)
    \,,
\end{eqnarray}
and $S_u$ preserves the scalar product, i.e.\ is a Lorentz transformation.
Finally
\begin{equation}\label{28I22.16bcd}
 (S_u)^2(\Ytov):= - u \cdot  S_u (\Ytov)  \cdot u ^{-1}
 =   u \cdot u \cdot \Ytov \cdot u^{-1} \cdot u ^{-1}
 =   \eta(u,u) \cdot \Ytov \cdot \frac{1}{\eta(u,u)}
 =
 \Ytov
 \,,
 \end{equation}
so $(S_u)^2$ is the identity map.

We continue with a  proof of \eqref{9I25.1}, namely
\begin{equation}\label{10I25.6}
S_{u+v} (u )= - v
\,,
\end{equation}
assuming that
\begin{eqnarray}\label{10I25.5a}
  &
  \eta(u,u)=\eta(v,v) \,, \ \mbox{and }
  \\
  &
  \eta(u+v,u+v)  \ne 0
   \,.
   \label{10I25.5b}
\end{eqnarray}
Equation \eqref{10I25.5a} gives
\begin{align}\label{10I25.6b}
  u\cdot u = v\cdot v
  \
  \Rightarrow
   \quad
 (u+v)\cdot u  =  &
 \
    u \cdot u +v\cdot u =
    v \cdot v +v\cdot u =
    v \cdot (v +  u )
    \nonumber
\\ = & \
    v \cdot (u +  v )
    \,.
\end{align}
Multiplying by $(u+v)^{-1}$ from the right we obtain the desired equation \eqref{10I25.6}:
\begin{align}\label{10I25.7a}
 S_{u+v} ( u)   =  &
 \
  - (u+v)\cdot u  \cdot (u +  v ) ^{-1}  =
   -  v
    \,.
\end{align}
This implies in particular that $S_{u+v} S_u$ maps $u$ into $v$:
\begin{align}\label{10I25.7b}
 S_{u+v} S_u ( u)   =  &
   S_{u+v}  (- u)   =
 - S_{u+v}  (  u)   =
     v
    \,.
\end{align}

We continue by noting that  conjugation by a product $u\cdot v$ is the product of   conjugations. Indeed, using associativity,
\begin{equation}\label{10I25.8}
  (u \cdot v )\cdot w \cdot (u \cdot v)^{-1}
  = u \cdot v \cdot w \cdot v ^{-1}\cdot u^{-1}  = S_u S_v (w)
  \,.
\end{equation}
So, assuming again \eqref{10I25.5a}-\eqref{10I25.5b}, the product property together with \eqref{10I25.6b} gives
\begin{equation}\label{10I25.9}
  S_{u+v} S_u   =  S_{v} S_{u+v}
  \,.
\end{equation}
Using this   we obtain
\begin{eqnarray}
  (S_{u+v} S_u) ^2
    & =  & \underbrace{S_{u+v} S_u}_{S_{v} S_{u+v}} S_{u+v} S_u
    \nonumber
\\
    & =  &  S_{v} \underbrace{ S_{u+v}  S_{u+v}}_{\Id} S_u
    \nonumber
\\
   &  = & S_{v} S_{u}
  \,.\label{10I25.9b}
\end{eqnarray}
This provides  a rather simpler derivation of the formulae underlying the claims of
 Proposition~\ref{p28I22.2}.

\section{Boosts and rotations via Clifford algebras}

Points 2.\ and 3.\ of Proposition  \ref{p28I22.2} together with Proposition \ref{p28I22.3}  give the following description of boosts and rotations:

\begin{prop}
 \label{p28I22.4}
 \begin{enumerate}
 \item
If $u$ and $v$ are unit, timelike, consistently time-oriented vectors   and if $u+v$ is not null, then
\begin{equation}\label{28I22.11b}
 \forall \ \Xtow\in \V  \qquad    B(v,u) \Xtow  = (u+v)\cdot u \cdot \Xtow  \cdot ( (u+v)\cdot u )^{-1}
  \,.
\end{equation}
\item
 If the  vectors $a $ and $b $ are linearly independent and orthogonal to a timelike vector $u$, then
 the map
 \begin{equation}\label{18I25.1}
  \V\ni \Xtow  \mapsto R(\Xtow )= a \cdot b \cdot \Xtow  \cdot (a\cdot  b)^{-1}
 \end{equation}
 is a rotation, in the
 oriented spacelike plane $\mathrm{Span}\{\red{a,b}\}$, by an  angle $-2\theta$, where $\theta$ is the (unoriented) angle between $a$ and $b$.
 \end{enumerate}
\end{prop}

\begin{Remark}
{\rm
The angle of rotation  in point 2.\ is negative because the map $R=S_a S_b$ defined by \eqref{18I25.1} is the inverse of the map $S_bS_a$ considered in Proposition~\ref{p28I22.2}.
}
\qed
\end{Remark}

Note that for any invertible element $c$ of the Clifford algebra we have
\begin{equation}\label{4II22.4}
 \forall \ \lambda \in \R^*\qquad  ( \lambda c )  \cdot \Xtow  \cdot (\lambda  c)^{-1}
   =c \cdot \Xtow  \cdot c^{-1}
   \,.
\end{equation}
 It follows that  the boost $B(v,u)$ is the conjugation by
$$
  \lambda (u+v)\cdot u = \lambda (-1 + v \cdot u )
 \,,
$$
and we are free to choose the real number $\lambda \ne 0$ as convenient. Similarly, for any $\sigma \in \R^*$
 the map $R$  in point 2.\ of  Proposition \ref{p28I22.4} is the conjugation by
\begin{eqnarray}
 \sigma\  a \cdot b
  & = &
 \sigma \big(\frac 12 (a \cdot b+ b \cdot a)
 +\frac 12
 ( a \cdot b -b \cdot a)
 \big)
  =   \sigma \big( \eta(a,b) +  \frac 12
  [a, b]
  \big)
 \nonumber
\\
 & = &   \sigma \big( |a| |b| \cos(\theta) +  \frac 12
  [a, b]
  \big)
   \label{422II.11}
  \,,
\end{eqnarray}
where the arbitrariness in the choice of $\sigma\ne 0 $ has been included for further use.

As the last preparatory step for the proof of~\eqref{4II22.2}
 we will need to identify the vectors defining the angle of the rotation out of formulae in the spirit of \eqref{422II.11}. This will be the contents of Proposition~\ref{l9II22.1} below, for the proof of which we will need the following Lemma:

\begin{lemma}
\label{l13I25.1}
Let $e_1$ and $e_2$ be unit vectors satisfying $e_1\perp e_2$ and let  $\mu, \nu \in \R$. If $\mu + \nu [e_1,e_2]=0$, then $\mu=\nu =0$.
\end{lemma}

\proof\
Since $e_1$ is orthogonal to $e_2$ we have $e_1 \cdot e_2=-e_2 \cdot e_1$, hence
\begin{equation}
 0 = \mu+ \nu  [e_1,e_2]  = \mu +\nu (e_1 \cdot e_2 - e_2 \cdot e_1) = \mu + 2 \nu e_1 \cdot  e_2
 \,.
 \label{13I25.21}
\end{equation}
Since $e_1\cdot e_1 =1$, multiplying \eqref{13I25.21} by $e_1$ gives
\begin{equation}
 0 =    \mu e_1  + 2 \nu    e_2
 \,,
 \label{13I25.22}
\end{equation}
which implies $\mu =\nu =0$, as desired.
\qedskip

We are ready now to prove:

\begin{proposition}
 \label{l9II22.1}
Let
\begin{equation}\label{WhenII22.3}
 r:= \alpha + \beta[p,q]
 \,,
\end{equation}
with $\alpha,\beta\in\R$, $ \beta  \ne 0$,
where
$p,q\in \V $ are spacelike, both orthogonal to a timelike vector, and linearly independent. Then $r$ is invertible,  and
  conjugation by $r$ defines
a Lorentz transformation $\Lambda$ which is a rotation  in the
oriented plane $\mathrm{Span}\{p,q\}$
 by an angle $-2\theta$  such that
\begin{equation}\label{9II22.11}
  \cot(\theta) =
     \frac{ \alpha}{2\beta(\eta(p,p) \eta(q,q) - \eta(p,q)^2)^{1/2} }
     \,,
     \quad   \
     {\redc \theta\in (0,\pi)}
  \,.
\end{equation}
\end{proposition}

\begin{remark}
  \label{R24I25.1}
  {\rm
  Let $0\ne a\in  \mathrm{Span}\{p,q\}$, then $\mathrm{Span}\{a,\Lambda a\}$ will have the same orientation as $   \mathrm{Span}\{p,q\}$ if  $\pi/2<\theta<\pi$, and opposite orientation if $0<\theta<\pi/2$.
  }
  \qed
\end{remark}

{\noindent\sc Proof:} 
We start by noting that if a vector $\Xtow $  is orthogonal to both $p$ and $q$, then it commutes with $[p,q]$. Indeed, by Proposition~\ref{P2II22.1} we have $\Xtow \cdot p = - p \cdot \Xtow $, similarly for $q$, hence
\begin{eqnarray*}
 \Xtow \cdot [p,q]
  &= &
    \Xtow  \cdot p \cdot q - \Xtow  \cdot q \cdot p
  = - p  \cdot \Xtow   \cdot q +  q \cdot \Xtow  \cdot p
  =     p \cdot q  \cdot \Xtow -  q \cdot p\cdot \Xtow
  \nonumber
\\
   &= &
    [p,q]\cdot \Xtow
  \,.
\end{eqnarray*}
This also shows that $\Xtow $ commutes with the Clifford element \eqref{WhenII22.3}. It follows that $\Lambda$ is the identity on $\mathrm{Span}\{p,q\}^\perp$, and hence a rotation of that plane.

The idea now is to find unit vectors
$a$ and $b$ in $\V$ so that $\Lambda $ is the conjugation by $a\cdot b$, hence \eqref{422II.11} applies  with some real number $\sigma\ne 0$:
\begin{equation}\label{9II22.1}
   \alpha + \beta [p,q]
    = \sigma
   \big(
    \cos(\theta) + \frac{1}{2}[a,b]
    \big)
  \,.
\end{equation}
For this let $\{e_1,e_2\}$ be an ON basis of  $ \mathrm{Span}\{p,q\}$  such that the transition matrix from $\{e_1,e_2\}$ to $\{p,q\}$ has positive determinant. We choose $a:=e_1$ and  write
$$
 b =
         \cos(\red{\theta})e_1 +
         \sin(\red{\theta})e_2
     \,,
     \quad
     p  = p_1 e_1 + p_2 e_2
     \,,
     \quad
    q   =q_1 e_1 + q_2  e_2
     \,,
$$
assuming that the desired vector $b$ exists.
 Then
\begin{eqnarray*}
 [a,b]
 &= &
  [e_1, \cos(\theta) e_1 + \sin(\theta) e_2] =
 \sin(\theta) [e_1,e_2]
 \,,
  \nonumber
\\
  {}[p,q]
   &= &
 [p_1 e_1 + p_2 e_2,q_1 e_1 + q_2 e_2]
 =
 (p_1 q_2 - p_2 q_1 ) [ e_1 , e_2]
 \,.
\end{eqnarray*}
Hence \eqref{9II22.1} can be rewritten as
\begin{eqnarray}
 \elltor
 \  \equiv
 \
  \alpha + \beta [p,q]
  & = &
 \alpha + \beta (p_1 q_2 - p_2 q_1 )  [ e_1 , e_2]
 \label{9II22.5a}
\\
  & = &  \sigma
   \big( \cos(\theta) + \frac{\sin(\theta)}2 [e_1,e_2]
    \big)
  \,,
   \label{9II22.5}
\end{eqnarray}
%


Before proving existence of $b$, let us show   that $\elltor$ is invertible. Indeed, we claim that
 \begin{align}\label{15I25.2}
   \elltor ^{-1} =   \sigma^{-1}
   \
   \big( \cos(\theta)- \ {\sin(\theta)} \,  e_1 \cdot e_2
    \big)
    \,.
 \end{align}
To show this we first    use
 that
 $$
  [e_1,e_2] = e_1 \cdot e_2- e_2 \cdot  e_1  = 2 e_1 \cdot e_2
  \,,
 $$
 as $e_1$ and $e_2$ are orthogonal. Thus
 \begin{align}\label{15I25.1}
   \elltor  = \sigma
   \big( \cos(\theta)+\frac{\sin(\theta)}2 [e_1,e_2]
    \big)  &   =
    \sigma   \big( \cos(\theta)+   {\sin(\theta)} \   e_1 \cdot e_2
    \big) \,,
 \end{align}
and we have
 \begin{align}\label{15I25.3}
   \elltor \cdot \sigma^{-1}
   \cdot
   &
   \big( \cos(\theta)-   {\sin(\theta)} \  e_1 \cdot e_2
    \big)
    \nonumber
\\
   = & \  \big( \cos(\theta)+   {\sin(\theta)}\   e_1 \cdot e_2
   \big)\cdot \big( \cos(\theta)-   {\sin(\theta)} \   e_1 \cdot e_2
   \big)
    \nonumber
\\
   = &  \  \cos^2 (\theta)  -  {\sin^2 (\theta)} \
   \underbrace{e_1 \cdot e_2
   \cdot   e_1 \cdot e_2}_{= - e_1 \cdot e_1
   \cdot   e_2 \cdot e_2 = -1\cdot 1 = -1
   }
    \nonumber
\\
   = & \   \cos^2 (\theta)  +   {\sin^2 (\theta)}  =1
   \,,
 \end{align}
which is the same as \eqref{15I25.2}.

In view of Lemma~\ref{l13I25.1}, a comparison of coefficients in \eqref{9II22.5} gives
\begin{eqnarray}
 \alpha = \sigma
    \cos(\theta)
    \,,
    \quad  \beta (p_1 q_2 - p_2 q_1 )  =  \sigma  \frac{\sin(\theta)}2
  \,.
   \label{9II22.5xd}
\end{eqnarray}
Eliminating $\sigma$ we obtain
\begin{equation}\label{9II22.6}
  \cot(\theta) = \frac{ \alpha}{2\beta (p_1 q_2 - p_2 q_1)}
  \,,
\end{equation}
which  together with the requirement that $\theta\in(0,\pi)$ determines  $\cos(\theta)$ up to a sign,  and establishes existence of the desired pair $(a,b)$ after choosing the sign of $\sigma$ so that  \eqref{9II22.5xd} holds.

To finish the proof we need to rewrite \eqref{9II22.6} in a basis-independent form. For this note first
that $p_1 q_2 - p_2 q_1$ is the determinant of the matrix
$$
 A:= \left(
       \begin{array}{cc}
        p_1 &  q_1 \\
        p_2 &  q_2
       \end{array}
     \right)
     \,.
$$
Now, the vectors $\{f_1,f_2\}:= \{p,q\}$  form, by definition, a positively oriented basis of $\mathrm{Span}\{p,q\} $. Let $\eta(f_a,f_b)$ be the $2\times2  $ matrix of their scalar products. By definition of the matrix $A$ we have $f_a = A^b{}_a e_b$, thus
\begin{equation}\label{9II22.4}
  \eta(f_a,f_b) = A^c{}_a A^d{}_b \eta(e_c,e_d)
  \,.
\end{equation}
Note that the matrix of the scalar products $\eta(e_c,e_d)$ is the identity matrix. Equation~\eqref{9II22.4} can be viewed as a matrix equation; taking determinants of both sides one then obtains
%
\begin{equation}\label{9II22.7}
  \det ( \eta(f_a,f_b)) = (\det A)^2 \det ( \eta(e_a,e_b)) = (\det A)^2
  \,.
\end{equation}
Equivalently,
\begin{equation}\label{9II22.8}
   \left|
       \begin{array}{cc}
        \eta(p,p) &  \eta(q,p) \\
        \eta(p,q) &   \eta(q,q )
       \end{array}
     \right|  = (p_1 q_2 - p_2 q_1)^2
     \,.
\end{equation}
This, together with \eqref{9II22.6}, proves \eqref{9II22.11}.
\qedskip

\section{Proof of \eqref{4II22.2}}
 \label{s5II22.5}

We are ready now to address the problem at hand, namely properties of the map
$$R= B(u,w) B(w,v) B(v,u)
$$
with three unit timelike vectors $u$, $v$ and $w$ such  that no pairwise sum is null. It follows from Proposition \ref{p28I22.4} that    $R$ is the  conjugation by
\begin{eqnarray}
 r
  &
 :=
  &
  (-1 + u \cdot w)  \cdot  (-1 + w \cdot  v) \cdot  (-1 +v \cdot u )
  \nonumber
\\
    &=&
  (-1 + u \cdot w)  \cdot   (  1 -  w \cdot  v  - v\cdot u +  w \underbrace{\cdot v \cdot v}_{-1} \cdot u
  )\nonumber
\\
    &=&
   -   (  1 -  w \cdot  v  - v\cdot u -   w   \cdot u
  )
   + u \cdot w     -   u \cdot w \cdot w \cdot  v  - u \cdot w  \cdot v\cdot u
    \nonumber
\\
 &&    -   u \cdot w \cdot  w   \cdot u
  \nonumber
\\
    &=&
   -1  +   w \cdot  v   +  v\cdot u  +
   \underbrace{ w   \cdot u
   + u \cdot w }_{2\, \eta (u,w)}    +   u   \cdot  v  - u \cdot w  \cdot v \cdot u -   1 \nonumber
\\
    &=&
   -2  +   \underbrace{w \cdot  v}_{2\, \eta(w,v) - v \cdot w}   +  2\eta (v,  u)   +   2\eta( w, u)
       - u \cdot w  \cdot v \cdot u\nonumber
\\
    &=&
   -2  +  2\, \eta(w,v)  +  2\eta (v,  u)   +   2\eta( w, u) - v \cdot w
       - u \cdot w  \cdot v \cdot u
    \,.
     \label{29I22.11}
\end{eqnarray}

The remaining Clifford-products in \eqref{29I22.11} are best analysed using the decomposition of $w$ and $v$ into parts parallel and orthogonal to $u$, as in \eqref{2II21.1}. By Proposition~\ref{P2II22.1} the vectors $\red{w_\parallel}$,  $\red{v_\parallel}$ and  $u$ all commute,
while  $\red{w_\perp}$ and $\red{v_\perp}$  anticommute with all three $\red{w_\parallel}$,  $\red{v_\parallel}$ and  $u$. This leads to
\begin{eqnarray}
v   \cdot w  & = &
  (\red{v_\parallel}   +  \red{v_\perp} )  \cdot    ( \red{w_\parallel} +
         \red{w_\perp} )
   \nonumber
\\
   & = &
   \red{v_\parallel}     \cdot    \red{w_\parallel} +
         \red{v_\parallel}     \cdot  \red{w_\perp}
         +
    \red{v_\perp}    \cdot     \red{w_\parallel} +
       \red{v_\perp}    \cdot   \red{w_\perp}
       \,,
          \label{2II22.22}
\\
   u \cdot w  \cdot v \cdot u
  & = &
     u \cdot (\red{w_\parallel} + \red{w_\perp})  \cdot   v \cdot u
   \nonumber
\\
  & = &
      (\red{w_\parallel} -  \red{w_\perp})  \cdot  u   \cdot   v \cdot u
   \nonumber
\\
  & = &
       (\red{w_\parallel} -  \red{w_\perp})  \cdot  \underbrace{ u   \cdot   u}_{-1}  \cdot  ( \red{v_\parallel} -
         \red{v_\perp} )
   \nonumber
\\
  & = &
      - (\red{w_\parallel}   -  \red{w_\perp} )  \cdot    ( \red{v_\parallel} -
         \red{v_\perp} )
   \nonumber
\\
  & = &
      - \red{w_\parallel}          \red{v_\parallel}
       +  \red{w_\parallel}    \cdot   \red{v_\perp}
      +   \red{w_\perp}    \cdot      \red{v_\parallel}
      -
          \red{w_\perp}    \cdot    \red{v_\perp} \nonumber
\\
  & = &
      - \red{w_\parallel}          \red{v_\parallel}
       -  \red{v_\perp}    \cdot     \red{w_\parallel}
      -    \red{v_\parallel}    \cdot   \red{w_\perp}
      -
          \red{w_\perp}    \cdot    \red{v_\perp}
    \,.
          \label{2II22.23}
\end{eqnarray}
Adding, one obtains
\begin{eqnarray}
v   \cdot w  +  u \cdot w  \cdot v \cdot u
   & = &
          \red{v_\perp}    \cdot   \red{w_\perp}
      -
          \red{w_\perp}    \cdot    \red{v_\perp}  \nonumber
\\
  & = & [\red{v_\perp}   ,  \red{w_\perp} ]
    \,.
          \label{2II22.23b}
\end{eqnarray}
Inserting into \eqref{29I22.11}   we find that
$$
 R(\Xtou ) = r \Xtou  r^{-1}
\,,
$$
%
where
\begin{eqnarray}
  r &= &
   -2
   \Big( 1      -   \eta (v,  u)   -    \eta( w, u)
   -    \eta(w,   v)
+ \frac 12 [\red{v_\perp},\red{w_\perp} ]
\Big)
    \,.
    \label{31I22.2}
\end{eqnarray}
We have thus brought $r$ to the form \eqref{WhenII22.3}, namely
\begin{equation}\label{WhenII22.3e}
 r= \alpha + \beta[p,q]
 \,,
\end{equation}
 with $\beta =-1$, $p=\red{v_\perp}$, $q=\red{w_\perp} $, and
\begin{equation}\label{16I25.41}
  \alpha=   -2
   \big( 1      -   \eta (v,  u)   -    \eta( w, u)
   -    \eta(w,   v)
   \big)
   \,.
\end{equation}
 Proposition~\ref{l9II22.1} shows that $R$ is a rotation of angle $-2\theta$ in the oriented plane $\mathrm{Span}\{\red{v_\perp},\red{w_\perp} \}$.

%

Recall \eqref{15I25.1}:
\begin{eqnarray}
 \elltor
  & = & \underbrace{ \sigma
   \big( \cos(\theta)
   }_{\alpha} + \sin(\theta) \  e_1\cdot e_2
    \big)
  \,.
   \label{9II22.5cd}
\end{eqnarray}
%
One could now use~\eqref{9II22.11} to obtain a formula for $\cot(\theta)$, and proceed from there.
It turns out that a simpler calculation proceeds by deriving  an explicit formula for $\sigma$, which together with  \eqref{16I25.41} will allow us to calculate $\cos(\theta)$. The formula can be obtained as follows.

Using $u\cdot u = -1$, $v\cdot v = -1$, etc., the definition~\eqref{29I22.11} of $r$ can be manipulated as
\begin{eqnarray}
 r
  &
 =
  &
  (-1 + u \cdot w)  \cdot  (-1 + w \cdot  v) \cdot  (-1 +v \cdot u )
  \nonumber
\\
    &=&
    u \cdot  (  u +  w)  \cdot w \cdot (w +    v) \cdot v\cdot  ( v + u )
    \,.
     \label{29I22.11dg}
\end{eqnarray}
Let us write $r^t$ for the element of the Clifford algebra where the order of all the products in $r$ has been reversed:
\begin{eqnarray}
 r^t
    &=&
    ( v + u ) \cdot v \cdot (w +      v)\cdot w\cdot  (  u +  w)   \cdot u
    \,.
     \label{29I22.11dg2}
\end{eqnarray}
After using
\begin{equation}\label{16I25.3}
   ( v + u )
    \cdot
    ( v + u )  =    v
    \cdot
     v + v\cdot  u   + u\cdot  v
   + u \cdot   u = -1 + 2 \eta(u,v) -1 = -2(1 + \gamma_{uv})
\end{equation}
one obtains
\begin{eqnarray}
|r|^2 &
 : =
  &   r \cdot r^t
  \nonumber
\\
  &
 =
  &
    u \cdot  (  u +  w)  \cdot w \cdot (w +     v) \cdot v\cdot
     \underbrace{( v + u )
    \cdot
    ( v + u )}_{-2(1 + \gamma_{uv})} \cdot v
    \nonumber
\\
 &&
   \phantom{xxxxxxxxxxxxxxxxxxxxxxxxxxxxx}\cdot  (w +    v)\cdot w\cdot  (  u +  w)   \cdot u
   \nonumber
\\
  &
 =
  &
    -2(1 + \gamma_{uv})\ u \cdot  (  u +  w)  \cdot w \cdot
    \underbrace{(w +     v) \cdot \underbrace{ v
     \cdot v }_{-1}\cdot  (w +    v)
     }_{-2(1+\gamma_{vw})}
     \cdot w\cdot  (  u +  w)   \cdot u
      \nonumber
\\
  &
 =
  &
    -2^2 (1 + \gamma_{uv})(1+\gamma_{vw})\ u \cdot  (  u +  w)  \cdot w
     \cdot w\cdot  (  u +  w)   \cdot u
  \nonumber
\\
  &
 =
  &
    2^3 (1 + \gamma_{uv})   (1 + \gamma_{vw})    (1 + \gamma_{wu})
    \,.
     \label{29I22.11dgr}
\end{eqnarray}

To continue, we rewrite $r^t$ as follows:
\begin{eqnarray}
 r^t
    &=&
    \underbrace{( v + u ) \cdot v} \cdot \underbrace{(w +      v)\cdot w}\cdot
      \underbrace{(  u +  w)   \cdot u }
      \nonumber
\\
    &=&
    ( -1 + u   \cdot v)  \cdot  (-1  +      v \cdot w ) \cdot
      (- 1 +  w   \cdot u)
    \,.
     \label{29I22.11dg24}
\end{eqnarray}
Comparing with the first line of \eqref{29I22.11dg}, namely
\begin{eqnarray}
 r
  &
 =
  &
  (-1 + u \cdot w)  \cdot  (-1 + w \cdot  v) \cdot  (-1 +v \cdot u )
    \,,
     \label{29I22.11dg5}
\end{eqnarray}
we see that $r^t$ differs from $r$ by interchanging $w$ with $v$. So the calculation leading to \eqref{31I22.2} applies and gives
\begin{eqnarray}
  r^t &= &
   -2
   \Big( 1      -   \eta (v,  u)   -    \eta( w, u)
   -    \eta(w,   v)
- \frac 12 [\red{v_\perp},\red{w_\perp} ]
\Big)
    \,.
    \label{31I22.2-}
\end{eqnarray}
Likewise \eqref{9II22.5cd} holds with the sign of the commutator term reversed:
\begin{eqnarray}
 \elltor ^t
  & = &
    \sigma   \big( \cos(\theta) -   {\sin(\theta)} \   e_1 \cdot e_2
    \big)
  \,.
   \label{9II22.5cdc}
\end{eqnarray}
A calculation identical to \eqref{15I25.3}
gives
 \begin{align}\label{15I25.356}
   \elltor \cdot
   \elltor^t
   = & \   \sigma^2  \big( \cos(\theta)+   {\sin(\theta)}\   e_1 \cdot e_2
   \big)\cdot \big( \cos(\theta)-   {\sin(\theta)} \   e_1 \cdot e_2
   \big)
    \nonumber
\\
   = &\   \sigma^2  \big(  \cos^2 (\theta)  -  {\sin^2 (\theta)} \
   \underbrace{e_1 \cdot e_2
   \cdot   e_1 \cdot e_2}_{= - e_1 \cdot e_1
   \cdot   e_2 \cdot e_2 = -1\cdot 1 = -1
   }
   \big)
    \nonumber
\\
   = & \  \sigma^2
   \,.
 \end{align}
Using \eqref{29I22.11dgr} we conclude that
\begin{eqnarray}
\sigma^2
  &
 =
  &
    2^3 (1 + \gamma_{uv})   (1 + \gamma_{vw})    (1 + \gamma_{wu})
    \,.
     \label{29I22.11dgrc}
\end{eqnarray}
Using \eqref{16I25.41}-\eqref{9II22.5cd}  we obtain
\begin{equation}\label{16I25.7}
  \cos^2 (\theta) = \frac{\alpha^2}{\sigma^2}
   = \frac{ 4
   \big( 1      +\gamma_{uv}    +\gamma_{vw} +\gamma_{wu }
   \big)^2}{2^3 (1 + \gamma_{uv})   (1 + \gamma_{vw})    (1 + \gamma_{wu})  }
   \,.
\end{equation}
Taking  orientation into account, by Proposition~\ref{l9II22.1} the    angle $\psi$ of the Thomas-Wigner rotation  is twice the angle $\theta$ in the oriented plane  $\mathrm{Span}\{\red{w_\perp},\red{v_\perp} \}$. Thus
\begin{eqnarray}
  \cos(\psi) &=& \cos(2\theta)
  \ = \
   2 \cos^2(\theta) - 1
   \nonumber
    \\
    &=& \frac{
   \big( 1      +\gamma_{uv}    +\gamma_{vw} +\gamma_{wu }
   \big)^2}{  (1 + \gamma_{uv})   (1 + \gamma_{vw})    (1 + \gamma_{wu})  }
   -1
   \,,
\end{eqnarray}
which is the desired Macfarlane's formula \eqref{4II22.2}.



\appendix

\section{An example of Clifford algebra of $(\R^3,\eta)$}
 \label{App30I22.1}

An example of a set of four-by-four \emph{real-valued} matrices $\gamma_\mu$ satisfying the Clifford relations for the three dimensional Mikowski scalar product, in the mostly-plus signature,  is
\begin{equation}\label{28I22.14}
   \gamma_0=  \left(
                \begin{array}{cc}
                  0 & \mathbbm{1}_{ 2\times 2} \\
                  -\mathbbm{1}_{ 2\times 2}  & 0 \\
                \end{array}
              \right)
 \,,
 \qquad \gamma_1=  \left(
                \begin{array}{cc}
                   \sigma_1 & 0 \\
                  0 & -  \sigma_1 \\
                \end{array}
              \right)
 \,,
 \qquad \gamma_2=  \left(
                \begin{array}{cc}
                    \sigma_3 & 0 \\
                  0 & - \sigma_3 \\
                \end{array}
              \right)
 \,,
\end{equation}
where $\mathbbm{1}_{ 2\times 2} $ is the two-by-two identity matrix, while the $\sigma_i$'s are the  Pauli matrices
\begin{equation}\label{28I22.18}
   \sigma_1=  \left(
                \begin{array}{cc}
                  0 & 1 \\
                  1  & 0 \\
                \end{array}
              \right)
 \,,
 \qquad \sigma_3=  \left(
                \begin{array}{cc}
                   1 & 0 \\
                  0 & -1 \\
                \end{array}
              \right)
 \,.
\end{equation}

\section{Proof of point 1.\ of Proposition~\protect\ref{P2II22.1}}
 \label{App7II22.1}

As already pointed out, if $\Xtou =0$ then the commutator $[\Xtou ,\Ytov  ]=[0,\Ytov  ]$ vanishes, while  $0=\Xtou =0 \times  \Ytov  $ which means that $\Xtou $ and $\Ytov  $ are collinear, and our claim holds in this case. It remains to consider the case $\Xtou \ne 0$.

Suppose then that one of the vectors, say $\Xtou $, is non-zero and    \emph{not null}. We  can then decompose $\Ytov  $ as in \eqref{2II22.5}, leading to
\begin{eqnarray*}
 0 &= & [\Xtou ,\Ytov  ] = [\Xtou ,\red{\Ytov_\parallel} + \red{\Ytov_\perp}] =
  \underbrace{[\Xtou ,\red{\Ytov_\parallel}  ]}_0 + [\Xtou , \red{\Ytov_\perp}]
 \nonumber
\\
  &   = &
    \Xtou  \cdot  \red{\Ytov_\perp} -
  \underbrace{ \red{\Ytov_\perp} \cdot  \Xtou  }_{=-  \red{\Ytov_\perp} \cdot  \Xtou  \ \mbox{since $\eta(\red{\Ytov_\perp},\Xtou )=0$}}
 \nonumber
\\
 & = &
  2 \Xtou  \cdot \red{\Ytov_\perp}
  \,.
\end{eqnarray*}
Thus   $ \Xtou  \cdot \red{\Ytov_\perp} $   vanishes. Then
\begin{equation}\label{7II22.1}
  0=  \Xtou \cdot  \Xtou
    \cdot \red{\Ytov_\perp} = \eta(u,u) \Ytov_\perp
     \,,
\end{equation}
and since $\eta(u,u)\ne 0$ we find that $\Ytov_\perp $ vanishes, and $\Ytov  $ is collinear with $\Xtou $.

The case, where $\Ytov  $ is \emph{not}  null or vanishes while  $\Xtou $ is null, follows from what has been proved so far by renaming $\Ytov  $ to $\Xtou $ and $\Xtou $ to $\Ytov  $.

It remains to consider the possibility that
both $\Xtou $ and $\Ytov  $ are null and commute. Then either $\Ytov  $ or $-\Ytov  $ has the same time-orientation as $\Xtou $. Replacing $\Ytov  $ by $-\Ytov  $ if necessary we can therefore assume that $\Xtou $ and $\Ytov  $ are null, consistently time oriented, and commute. If $\Ytov  $ is not collinear with $\Xtou $ then  $\Xtou +\Ytov  $ is timelike and commutes with $\Ytov  $. From what we just proved $\Ytov  $ is then collinear with $\Xtou +\Ytov  $, hence timelike, a contradiction. Thus $\Xtou $ and $\Ytov  $ are also collinear in this case, and the proof is complete.
\qedskip
%

%
%

\noindent {\sc Acknowledgements:} We are grateful to Thomas Mieling for useful comments about a previous version of this manuscript.
\providecommand{\bysame}{\leavevmode\hbox to3em{\hrulefill}\thinspace}
\providecommand{\MR}{\relax\ifhmode\unskip\space\fi MR }
\providecommand{\MRhref}[2]{%
  \href{http://www.ams.org/mathscinet-getitem?mr=#1}{#2}
}
\providecommand{\href}[2]{#2}


\begin{thebibliography}{1}

\bibitem{Berger}
M.~Berger, \emph{G\'eom\'etrie}, Nathan, Paris, 1990.

\bibitem{Macfarlane}
A.J. Macfarlane, \emph{On the restricted Lorentz group and groups
  homomorphically related to it}, Jour.\ Math.\ Phys. \textbf{3} (1962),
  1116--1129.

\bibitem{SilbersteinBook}
L.~Silberstein, \emph{The theory of relativity}, MacMillan, London, 1914.

\bibitem{ThomasPrecession}
L.H. Thomas, \emph{The motion of the spinning electron}, Nature \textbf{117}
  (1926), 514.

\bibitem{UrbantkeThomas}
H.K. Urbantke, \emph{{Physical holonomy, Thomas precession, and Clifford
  algebra}}, Am.\ Jour. Phys. \textbf{58} (1990), 747--750.

\bibitem{WignerPrecession}
E.~Wigner, \emph{On unitary representations of the inhomogeneous Lorentz
  group}, Annals of Math. \textbf{40} (1939), 149--204.

\end{thebibliography}
\end{document}